\documentclass[showpacs,prb,twocolumn,amsmath,amssymb,floatfix]{revtex4-1}
\usepackage{graphicx}
\usepackage{dcolumn}
\usepackage{bm}
\usepackage{color}

\begin{document}

\title{Magnetization dynamics and frustration in the multiferroic double perovskite Lu$_2$MnCoO$_6$}

\author{Vivien S. Zapf,$^1$ B. G. Ueland,$^2$ Mark Laver,$^3$ Martin Lonsky,$^4$ Merlin Pohlit,$^4$ Jens M\"{u}ller,$^4$ Tom Lancaster,$^5$ Johannes S. M\"oller,$^{6,*}$ Stephen J. Blundell,$^6$ John Singleton,$^1$  Jorge Mira,$^7$ Susana Ya\~{n}ez-Vilar,$^8$ and Maria Antonia Se\~{n}ar\'{i}s-Rodr\'{i}guez$^8$}

\affiliation{$^1$National High Magnetic Field Laboratory, Los Alamos National Laboratory, Los Alamos, NM 87545, USA}
\affiliation{$^2$Ames Laboratory, U.S. DOE and Dept. of Physics, Iowa State University, Ames, Iowa 5011, USA}
\affiliation{$^3$School of Metallurgy and Materials, University of Birmingham, Edgbaston B15 2TT, UK} 
\affiliation{$^4$Institute of Physics, Goethe University, 60438 Frankfurt (M), Germany}
\affiliation{$^5$Durham University, Department of Physics South Road, Durham DH1 3LE, UK}
\affiliation{$^6$Oxford University, Department of Physics, Clarendon Laboratory, Oxford OX1 3PU, UK}
\affiliation{$^7$Departamento de F\'{i}sica Aplicada, Universidad de Santiago de Compostela, E-15782 Santiago de Compostela, Spain}
\affiliation{$^8$Dpto. Qu\'{i}mica Fundamental U. Coru\~{n}a, 15071 A Coru\~{n}a, Spain}
\affiliation{$^*$Now at the Laboratory for Solid State Physics, ETH Z\"urich, Z\"urich, Switzerland}

\begin{abstract}

We investigate the magnetic ordering and the magnetization dynamics (from kHz to THz time scales) of the double perovskite Lu$_2$MnCoO$_6$ using elastic neutron diffraction, muon spin relaxation and micro-Hall magnetization measurements. 
This compound  is known to be  a type II multiferroic with the interesting feature that a ferromagnetic-like magnetization hysteresis loop couples to an equally hysteretic electric polarization in the bulk of the material despite a zero-field magnetic ordering of the type $\uparrow \uparrow \downarrow \downarrow$  along Co-Mn spin chains. Here we explore the unusual dynamics of this compound and find extremely strong fluctuations, consistent with the axial next-nearest-neighbor Ising  (ANNNI) model for frustrated spin chains. We identify three temperature scales in Lu$_2$MnCoO$_6$ corresponding to the onset of highly fluctuating long-range order below $T_N = 50 \pm 3$ K identified from neutron scattering, the onset of magnetic and electric hysteresis, with change in kHz magnetic and electric dynamics below a 30 K temperature scale, and partial freezing of $\sim$MHz spin fluctuations in the muon spin relaxation data below $12 \pm 3$ K.  Our results provide a framework for understanding the multiferroic behavior of this compound and  its hysteresis and dynamics.
\end{abstract}

\pacs{Valid PACS appear here}

\maketitle
Magnetic order that induces electric polarization is a focus area of multiferroic research. This cross coupling of magnetism and ferroelectricity involves intriguing physics and is motivated by applications in electronics, sensing and electronic memory. \cite{Eerenstein06,Scott07,Nan08,Garcia10,Wu10,Chen11,Sun12} However, finding a material where a net magnetization $M$ (e.g. not a purely antiferromagnetic order) couples strongly to electric polarization $P$ is rare, particularly outside of heterostructures. Multiferroics are often divided into two categories. \cite{Khomskii09} In type I multiferroics, the magnetic and electric order parameters are distinct from each other with different ordering temperature so only a small fraction of $M$ and $P$ couple to each other, often via lattice strain. This phenomenology can occur in bulk materials, or in heterostructures where one material is ferromagnetic and the other ferroelectric and coupling occurs at an interface. In type II multiferroics on the other hand,  $P$ is entirely induced by magnetic order. A feature of type II multiferroics is that the entire electric polarization can be switched by magnetic field, or magnetization by electric field. However, type II multiferroics usually require a magnetic order that breaks spatial-inversion symmetry of the spin-lattice system and such orderings typically have very little net ferromagnetic-like magnetization.

Here we study the type II multiferroic compound Lu$_2$MnCoO$_6$, \cite{YanezVilar11,Lee14} where a net hysteretic $M$ couples to a net hysteretic $P$. It forms in the double perovskite A$_2$BB'O$_6$ structure with a slight monoclinic distortion.  Magnetic order has been reported below $T_N = 43 $K (which we here correct to be 50 K) and remarkably, significant hysteresis loops are observed in applied magnetic fields in both the magnetization and ferroelectric polarization below $\sim$ 30 K. \cite{YanezVilar11} 
The hysteretic magnetization in Lu$_2$MnCoO$_6$ is puzzling since it evolves out of an $H = 0$ magnetic order that has no net magnetization. The electric polarization $P$  is also hysteretic in applied magnetic fields in the sense that an electric polarization first established by cooling through $T_N$ in an electric field is irreversibly removed at the coercive magnetic field of the magnetization, resulting in a hysteretic $P(H)$. \cite{YanezVilar11} Previous elastic neutron scattering results in zero magnetic field, obtained at 4 K on a polycrystalline sample, have revealed $S = 3/2$ Co$^{2+}$ and Mn$^{4+}$ spins pointing along the $c$-axis, forming a pattern of ``$\uparrow$Co $\uparrow$Mn $\downarrow$Co $\downarrow$Mn",  or alternately ``$\uparrow$Mn $\uparrow$Co $\downarrow$Mn $\downarrow$ Co" that propagates along the $c$-axis. \cite{YanezVilar11} The neutron diffraction data also showed a long-wavelength modulation in the $ab$-plane, and the antiferromagnetic propagation vector was found to be  $\bm\tau = (0.0223(8),0.0098(7),0.5)$ at $H = 0$. \cite{YanezVilar11} This magnetic order in conjunction with the lattice does break spatial inversion symmetry and allows electric polarization to form. Unlike Ca$_3$CoMnO$_6$, \cite{Choi08} which also shows ``$\uparrow$Mn $\uparrow$Co $\downarrow$Mn $\downarrow$Co" ordering along chains, the electric polarization in Lu$_2$MnCoO$_6$ points along the $b$-axis, perpendicular to the $c$-axis Ising spins. \cite{Lee14,Chikara15}

We note that the $\uparrow \uparrow \downarrow \downarrow$ ordering is a ground state of the classic ANNNI (Axial Next-Nearest Neighbor Ising) model \cite{Bak82,Selke88} for magnetically frustrated Ising spin chains. This model has also been proposed to describe the magnetic behavior of Ca$_3$CoMnO$_6$. \cite{Choi08,Jo09,Kim14,Lancaster09,Kamiya12b} ANNNI models postulate Ising spin chains where nearest and next-nearest neighbor interactions within chains compete while interchain interactions are ferromagnetic. Moreover, some 3-D antiferromagnetic spin chain structures can be mapped on to a 1-D ANNNI scenario. \cite{Kamiya12b} In ANNNI models, many different close-lying magnetic states are predicted to occur as a function of small changes in external tuning parameters, with different incommensurate orderings. If the ANNNI scenario applies, it would provide a framework for understanding the unusual dynamics and hysteresis of the magnetic and resultant electric orders. Here we explore whether the magnetic behavior of Lu$_2$MnCoO$_6$ is consistent with a variant of the ANNNI scenario. In addition, we seek to answer questions about features in the magnetization and electric polarization as a function of temperature raised by previous works. \cite{YanezVilar11,Lee14} For example, is it truly a type II multiferroic if the magnetic order onsets below 50 K yet the electric hysteresis onsets below 30 K? Is there a second ordering transition at 30 K or does it correspond to a change in domain pinning? What is the origin of strong features in the magnetization that were observed at 12 K? 

In this work, we present results from temperature-dependent elastic neutron diffraction and muon spin relaxation ($\mu$SR) measurements on polycrystals, which probe the possibility of additional magnetic ordering transitions and investigate magnetic dynamics. We also show sensitive Hall magnetometry of sintered monocrystalline grains taken from the polycrystals. Our results are able to resolve the nature of the different features at 50, 30 and 12 K, multiple changes in dynamics, and strong fluctuations consistent with magnetic frustration.

\section{Experimental Details}
Elastic neutron scattering measurements were performed on polycrystalline samples at the Paul Scherrer Institut using the RITA II cold neutron triple axis spectrometer.  A pyrolytic graphite monochromator was used to select neutrons with wavelengths of $\lambda = 4.217$ \AA, and the collimation of the beam incident on the monochromator was limited by the neutron guide divergence ($m=2$).  A $40^{\prime}$  S\"{o}ller slit collimator was placed between the monochromator and sample, and a $180^{\prime}$ radial collimator was inserted between the sample and a multi-blade analyzer.  The multi-blade analyzer was tuned so that each blade selected a neutron energy transfer of $E=0$, and a two-dimensional position sensitive detector was employed.  In this configuration, each blade reflected neutrons corresponding to a different value of the scattering angle $2\theta$, and the difference in the $2\theta$ values corresponding to the two end blades was $\sim 5 ^{\circ}$. The multi-blade analyzer had an effective collimation of $\approx 40^{\prime}$.  Liquid nitrogen-cooled Be filters were placed before and after the sample to reduce contamination from higher-order neutron wavelengths.  The sample was cooled as low as $T=1.6$~K in a liquid He cryostat. The powdered sample was loaded into a vanadium can, and was the same one used in Ya\~{n}ez-Vilar {\it et al}. \cite{YanezVilar11} Comparison of measurements with and without methyl alcohol to freeze the power in place confirmed that the sub-micron crystalline powder was not moving during the measurements. 

$\mu$SR measurements were conducted at the Swiss Muon Source using the GPS spectrometer. The samples were packed inside 25 $\mu$m Ag foil and mounted on a $^{4}$He flow cryostat.

Magnetization was detected by Hall sensors on $\sim$40 monocrystalline grains of Lu$_2$MnCoO$_6$ with average diameter $\sim 1 - 2\,\mu$m extracted from the polycrystal. The grains were positioned on a Hall sensor with $10 \times 10\,\mu {\rm m}^2$ active area.  The individual crystallites show the faceted morphology that typically is displayed by single crystals of perovskites. The stray field emanating from the sample perpendicular to the Hall plane was recorded as $\langle B_z \rangle$ in a gradiometry configuration, by applying an amplitude and phase-adjusted current in opposite direction to an empty reference sensor. 
A small nonlinear background remains due to the ballistic nature of electron transport in the sensor and subtle differences between different Hall sensors. In addition, at $T = 0.3$ K for perpendicular orientations of the external field with respect to the sensor, quantum oscillations and the quantum Hall effect in the 2-D electron gas of the Hall sensor can be observed in the data. These non-hysteretic backgrounds are distinguishable from the magnetic hysteresis of the samples. Measurements were taken for $H$ parallel  ($H_{\parallel}$) and perpendicular ($H_{\perp}$) to the sensor plane in a temperature range from 50 K down to 0.3 K. The samples were attached to the Hall sensor due to surface forces and were observed not to move before and after measurement.

\section{Results}

\begin{figure}
\centering
\includegraphics[width=8 cm]{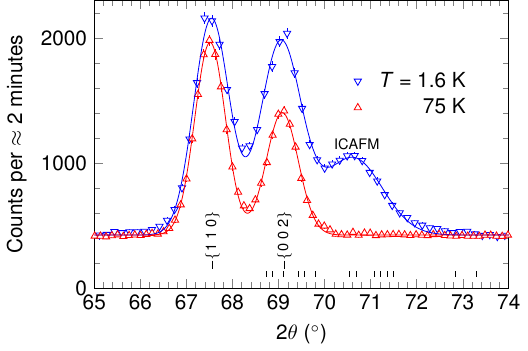}
\caption{Elastic neutron scattering data for $T = 1.6$ and 75~K. $\lambda = 4.217$ \AA.  The lines show Gaussian fits to the peaks. The ticks underneath the data indicate
the symmetry-allowed nuclear (top row) and magnetic (bottom row) Bragg positions. The \{1~1~0\} and \{0~0~2\} peaks are also labeled on the graph.  The broad peak labeled as ICAFM (incommensurate antiferromagnetism) contains contributions from multiple magnetic peaks. Table~\ref{Tab1} lists the magnetic Bragg peaks corresponding to the bottom row of tick marks.}
\label{ENS}
\end{figure}

 \begin{table}
 \caption{The symmetry-allowed magnetic Bragg peaks marked in Fig.~\ref{ENS} and their scattering angles for \mbox{$\lambda = 4.217$ \AA.}  The lattice parameters determined from fits to the neutron scattering data for $T=1.6$~K are $a=5.169(1)$~\AA, $b=5.559(1)$~\AA, $c=7.420(1)$~\AA, and $\beta=89.70(3)^{\circ}$.  }\label{Tab1}
 \begin{ruledtabular}
 \begin{tabular}{l @{\hspace{1.25cm}} r@{\hspace{-1.25cm}} r @{\hspace{-1.25cm}}l}
$2\theta$~($^{\circ}$) &  \multicolumn{3}{c}{\hspace{-1cm}Magnetic Bragg Peak}\\
\hline
68.730  &   $(1  -\tau_{\textrm{x}}$  &  $\pm1    \mp\tau_{\textrm{y}}$  &    $0       +\tau_{\textrm{z}})$ \\
68.875    &    $(-1 +\tau_{\textrm{x}}$ &     $\pm1    \mp\tau_{\textrm{y}}$     &     $1    -\tau_{\textrm{z}})$     \\
69.106   &    $(0  +\tau_{\textrm{x}}$ &     $\pm1    \mp\tau_{\textrm{y}}$     &     $1   +\tau_{\textrm{z}})$ \\
69.116   &    $(0  -\tau_{\textrm{x}}$ &    $\pm1    \mp\tau_{\textrm{y}}$  &    $1      +\tau_{\textrm{z}})$ \\
69.424  &    $(1  -\tau_{\textrm{x}}$ &    $\pm1    \pm\tau_{\textrm{y}}$  &    $0     +\tau_{\textrm{z}})$ \\
69.568    &    $(-1 +\tau_{\textrm{x}}$ &     $\pm1   \pm\tau_{\textrm{y}}$      &    $1    -\tau_{\textrm{z}})$     \\
69.798   &    $(0  +\tau_{\textrm{x}}$ &     $\pm1   \pm\tau_{\textrm{y}}$      &    $1    +\tau_{\textrm{z}})$ \\
69.808   &    $(0  -\tau_{\textrm{x}}$ &    $\pm1    \pm\tau_{\textrm{y}}$  &    $1    +\tau_{\textrm{z}})$ \\
70.535   &    $(1  +\tau_{\textrm{x}}$ &     $\pm1   \mp\tau_{\textrm{y}}$      &    $0    +\tau_{\textrm{z}})$ \\
70.685   &    $(1  -\tau_{\textrm{x}}$ &        $0    \pm\tau_{\textrm{y}}$  &    $1   +     \tau_{\textrm{z}})$ \\
71.074   &    $(-1 -\tau_{\textrm{x}}$ &    $\pm1    \mp\tau_{\textrm{y}}$  &    $1        -\tau_{\textrm{z}})$     \\
71.221    &    $(-1 +\tau_{\textrm{x}}$  &       $0    \pm\tau_{\textrm{y}}$      &    $1    +\tau_{\textrm{z}})$ \\
71.370   &    $(1  +\tau_{\textrm{x}}$ &     $\pm1\pm\tau_{\textrm{y}}$         & $0   + \tau_{\textrm{z}})$ \\
71.502   &    $(-1 -\tau_{\textrm{x}}$ &    $\pm1      \pm\tau_{\textrm{y}}$     &  $1       -\tau_{\textrm{z}})$     \\
72.847  &    $(1 +\tau_{\textrm{x}}$   &      $0    \pm\tau_{\textrm{y}}$           &$1  +  \tau_{\textrm{z}})$ \\
73.290   &    $(-1-\tau_{\textrm{x}}$  &      $0        \pm\tau_{\textrm{y}}$       &$1+        \tau_{\textrm{z}})$ \\
 \end{tabular}
 \end{ruledtabular}
 \end{table}

\begin{figure}
\centering
\includegraphics[width=8cm]{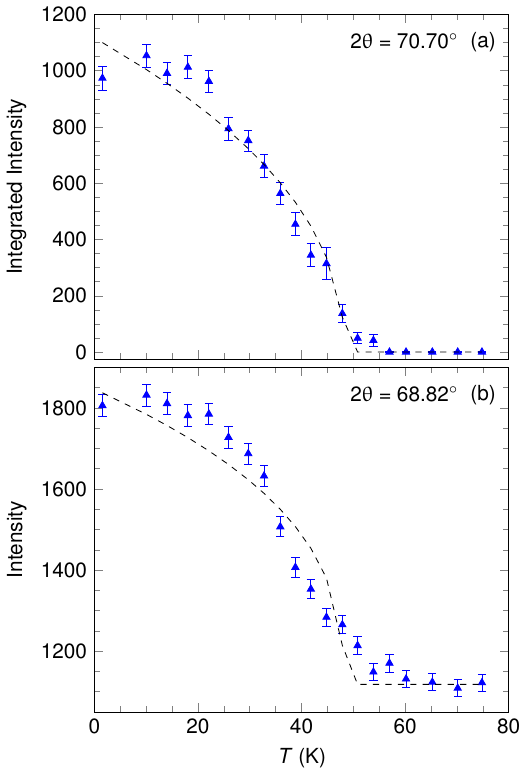}
\caption{The temperature dependence of the integrated intensity of the neutron diffraction peak, shown in Fig. $1$, near $70.70^{\circ}$ [ICAFM (a)] and the temperature dependence of the intensity of the elastic scattering at $68.82^{\circ}$ (b).  The integrated intensity of the 70.70$^{\circ}$ peak was determined from fits to a broad Gaussian lineshape. The dashed lines are guides to the eye, which have been drawn with $T_{\textrm{N}}=50$~K.}
\label{neutron_Tdep}
\end{figure}

Elastic neutron diffraction measurements at $T = 1.6$~K and 75 K are shown in Fig.~\ref{ENS} as a function of the scattering angle $2\theta$.  The top row of tick marks under the data shows the positions for the Bragg peaks due to the crystal structure, and the bottom row of tick marks indicates the positions of all of the symmetry-allowed Bragg peaks for the previously-determined magnetic order. \cite{YanezVilar11} The  magnetic Bragg peaks are also listed in Table~\ref{Tab1}.  Since the ordered moment lies along the $c$-axis \cite{YanezVilar11}, the increase in scattering near the \{0~0~2\} peak position is not due to an increase in the integrated intensity of the \{0~0~2\} peak with decreasing temperature.  Rather, it is due to an increase in the integrated intensities of the magnetic Bragg peaks lying close to the \{0~0~2\} position.  We refer to the broad scattering peak at $2\theta = 70.70^{\circ}$, which arises purely from magnetic Bragg peaks, as ICAFM (incommensurate antiferromagnetism).  The temperature dependence of the integrated intensity of the ICAFM peak and the intensity of the scattering for 68.82$^{\circ}$ are plotted in Fig.~\ref{neutron_Tdep}, which gives the temperature dependence of the magnetic order parameter.  The antiferromagnetic (AFM) transition occurs at $T_{\textrm{N}}=50 \pm 3$~K. 
An important point is that the data in Fig.~\ref{neutron_Tdep} indicate that long-range antiferromagnetic order occurs below $50$~K.  There may be a slight wiggle in the data in Fig.~\ref{neutron_Tdep}, near $40$~K, but taken as a whole, our current neutron scattering data do not show convincing evidence for a second transition occurring at $1.6$~K $\leq T \leq T_{\textrm{N}}$.

The ordering temperature from neutron scattering is slightly higher than the 43 K previously estimated from a peak in magnetization, \cite{YanezVilar11} indicating that the inflection point and not the peak of $M(T)$ corresponds to $T_N$.

\begin{figure}
\begin{center}
\includegraphics[width=8 cm]{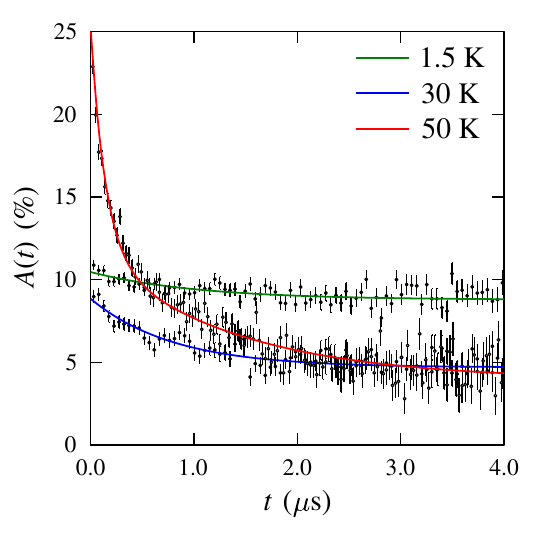}
\caption{Zero field $\mu^{+}$SR spectra showing the asymmetry in muon detection rates as a function of time, plotted here for temperatures of $T =$ 1.5, 30, and 50 K}. \label{muSR_spectra}
\end{center}
\end{figure}

Example  $\mu$SR spectra measured at three temperatures are shown in Fig.~\ref{muSR_spectra}, showing the asymmetry (proportional to the average muon polarization) as a function of time. An important point to note is that the spectra at all temperatures show monotonic relaxation with no oscillations. The spectra are typical of relaxation caused by dynamic fluctuations of the magnetic field distribution experienced by the muon ensemble. 

The behavior may be separated into three regimes. Above $T_N$ we observe relaxation with the full relaxing fraction of asymmetry. For $12 \leq T \leq 50$~K the initial asymmetry is reduced,
but the spectra continue to relax to the same baseline. For $T<12$~K the apparent baseline increases and the relaxation rate is reduced. 

\begin{figure}
\begin{center}
\includegraphics[width=8 cm]{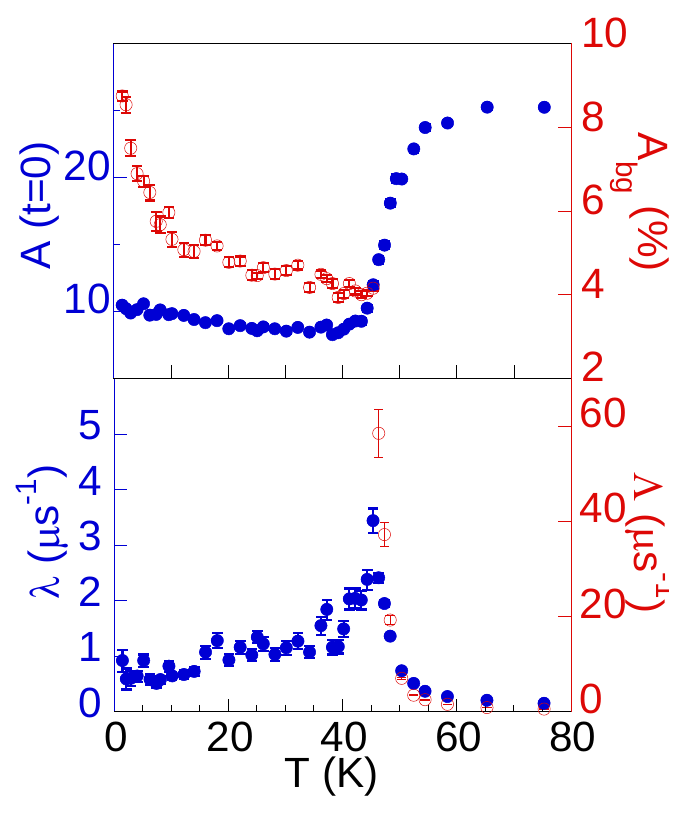}
\caption{Temperature dependences of the $\mu^{+}$SR asymmetry $A(t=0)$, and the quantities $\lambda$, $\Lambda$, and $A_\mathrm{bg}$ extracted from fits to the $\mu$SR spectra, as described in the text. }\label{muSR_Tdep}
\end{center}
\end{figure}

In order to parametrize the behavior of the system, the $\mu$SR spectra were fitted to two relaxation
functions. For $T>T_N$ the spectra were found to be well described by two exponential functions:
\begin{equation}
A(t) = A_{\lambda}e^{-\lambda t}+A_{\Lambda}e^{-\Lambda t}+A_{\mathrm{bg}},
\label{tgttn}
\end{equation}
where $\lambda$ and $\Lambda$ are relaxation rates. The two exponential functions imply the existence of two magnetically distinct
classes of muon site,  or, in the case of strongly Ising-like spins in a powder sample, the fact that 1/3 of the muons spins will initially be polarized parallel (or antiparallel) to the preferred local field direction and 2/3 will be
perpendicular, undergoing different relaxation processes as a result.

For $T<T_N$ we only resolve a single exponential relaxation rate and data are fitted to
\begin{equation}
A(t) = A_{\lambda}e^{-\lambda t}+A_{\mathrm{bg}}. 
\label{tlttn}
\end{equation}

The temperature dependence of $A_{\lambda}$, $\lambda$, $A_{\Lambda}$, $\Lambda$, and $A_\mathrm{bg}$ are shown in Fig.~\ref{muSR_Tdep}. We find that at temperatures well above $T_N = 50$~K, the amplitudes take constant values 
$A_{\lambda}=13.5$~\%, $A_{\Lambda}=7.7$\% and the baseline is $A_{\mathrm{bg}}=3.9$\%. 
We also find that the relaxation rate $\Lambda$ takes values far larger than $\lambda$, although the ratio of the two remains in a roughly fixed proportion, implying that they are tracking the same physics and only
differ due to the position of the respective muon site in the unit cell.  

The baseline amplitude $A_{\mathrm{bg}}$ is not constant below $T_N$, but rather increases significantly below
$T_f=12$~K. The evolution of the initial asymmetry $A_\lambda(t=0)$ with temperature shows a sharp drop around 50 K, which is typical of a system undergoing a magnetic ordering transition. 
The large, slowly fluctuating magnetic electronic moments that develop below $T_N$ cause muon spins to evolve
very rapidly upon implantation. The average polarization of these muons is dephased within the first time bin (1 ns) of the measurement and only a residual
relaxation is observed. However, the lack of oscillations in the asymmetry suggests that the transition is not one to a regime of
quasistatic long range magnetic order on the muon (microsecond) timescale. In this regime we expect the relaxation to
vary as $\lambda \propto \gamma_{\mu}^{2}\langle (B-\langle B\rangle)^{2}\rangle \tau$, where $\tau$ is the relaxation time, $B$ 
is the local magnetic field at the muon site(s) and $\gamma_{\mu}$ is the muon gyromagnetic ratio.

The large magnitude of  the relaxation could result from a broad distribution of static local magnetic fields or to dynamic fluctuations.  Of these two possibilities, the dynamic scenario is confirmed by the fact that the spectra in the region $12 \leq T \leq 50$~K relax to the same baseline value as for $T>T_N$.
 For a powder sample we would expect that 1/3 of muon spins should initially polarize along the direction of the magnetic field. In a {\it static} magnetic state these would not depolarize and would lead to an increase in the
apparent baseline of the relaxation compared to that found for $T>T_N$.  The state just below $T\approx 50$~K therefore appears to be one showing large moments with some degree of disorder, which are
dynamically fluctuating on the muon time-scale. 

The increase in the baseline amplitude $A_{\mathrm{bg}}$ below 12~K is strongly suggestive of a freezing out of these dynamics in the local
magnetic field distribution experienced by the muons. In this case the 2/3 of the muon spins initially oriented perpendicular to the local magnetic field direction will still be dephased by the variation in
the static fields across the muon ensemble, but those 1/3 of muon with spins directed parallel to the local magnetic fields will not be
dephased in the absence of dynamics. It is worth noting that this behavior was also observed in the low temperature behavior of Ca$_{3}$Co$_{2-x}$Mn$_{x}$O$_{6}$
for $x=0.95$ \cite{Lancaster09} and for other values of $x$. \cite{tom2}

Correlation times for the spin fluctuations are likely 10-100 ps. For example, a local magnetic field of 500~mT fluctuating with correlation times on the order of 10~ps would yield a relaxation rate on the order of 2~$\mu$s$^{-1}$, close to what we observe.

\begin{figure}
\centering
\includegraphics[width=8 cm]{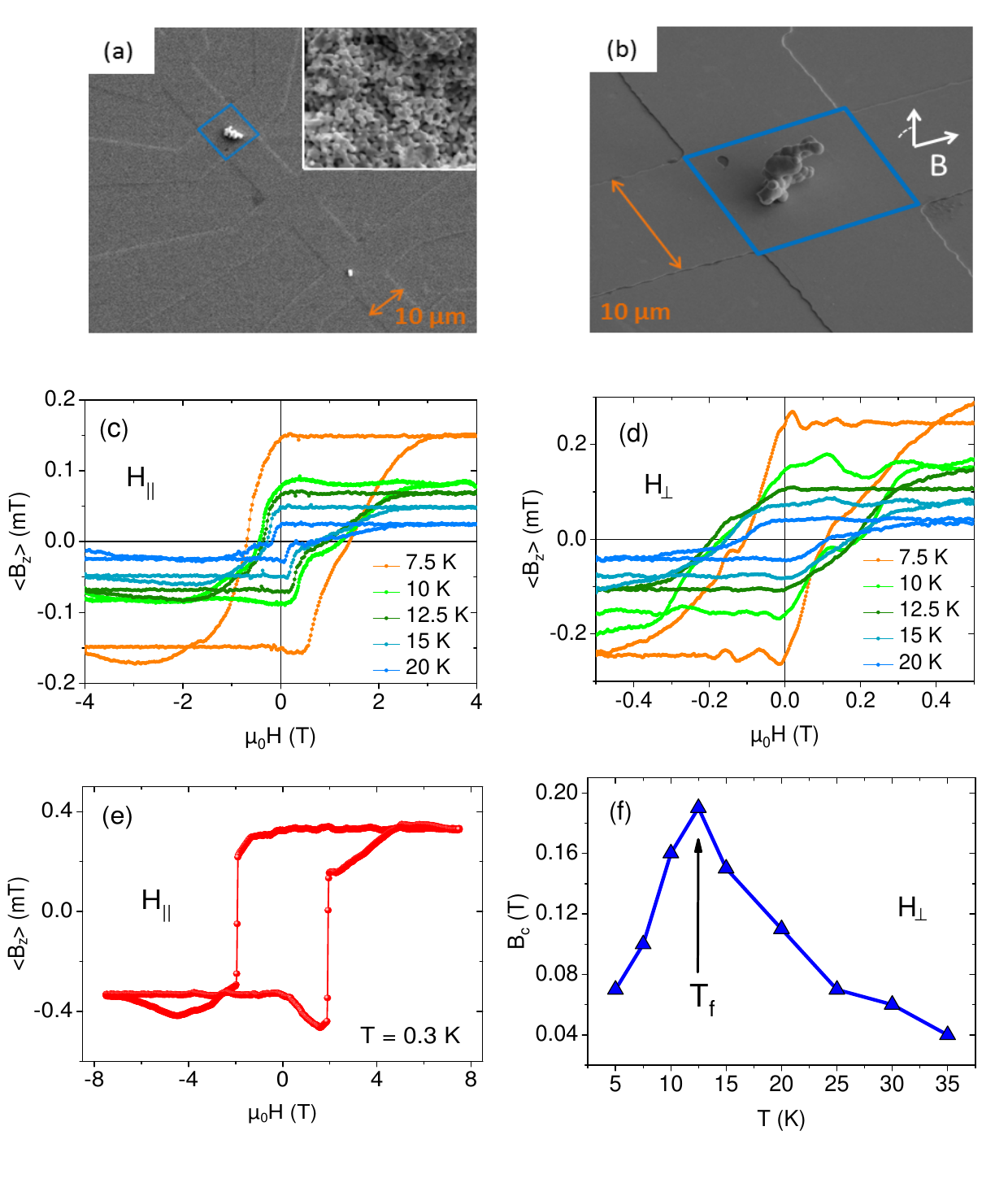}
\caption{ (a) Single crystal cluster positioned on a Hall sensor with $10\times10\,\mathrm{\mu m^{2}}$ active area  (blue square). The crystal cluster is extracted from a polycrystal shown in the inset. (b) Zoomed-in picture showing the sample morphology. (c)
Sample's stray field measured by the Hall sensor $\langle B_{\mathrm{z}} \rangle$  for $T$ between 7.5 and 20 K for $H_{\parallel}$ sensor plane (predominantly $H \parallel c$) and (d) for $H_{\perp}$ sensor plane (predominantly $H \perp c$). (e) Hysteresis loop at $T = 0.3$ K for $H_{\parallel}$. (f) Coercive magnetic field $B_c$ vs. temperature $T$ for $H_{\perp}$ the sensor plane, showing a peak that is labelled $T_f$.}
\label{MicroHall}
\end{figure}

Finally, we explore the temperature dependence of magnetic anisotropy in these polycrystals by micro-Hall measurements. Pictures of the collection of monocrystalline grains on the Hall sensor are shown in Figs.~\ref{MicroHall}(a)  and (b).
Fig.~\ref{MicroHall}(c), (d) and (e) shows magnetic hysteresis loops after background subtraction for applied $H$ parallel ($H_{\parallel}$) and perpendicular ($H_{\perp}$) to the sensor plane. 
We observe a strong anisotropy in  $M(H)$. Although due to the irregular sample shape a contribution from shape anisotropy cannot be excluded, in comparison to recent single crystal results \cite{Lee14}, we identify $H_{\parallel}$ as being predominantly  $H \parallel c$.
Unlike the single crystals where $H \perp c$ shows no hysteresis, we see a small hysteresis present for $H_{\perp}$, with a coercive magnetic field $H_c$ that is 10x smaller than for $H_{\parallel}$.   Exchange bias is observed for $H_{\parallel}$ but not $H_{\perp}$. 
 The hysteresis loop at $T = 0.3$ K reveals --  only for $H_{\parallel}$ -- a sharp jump in $M$ at a coercive field of 2 T, consistent with the studies of Y\'{a}\~{n}ez-Vilar {\it et al.}, where sudden switching has been observed at 2 T for $T \leq 2\,\mathrm{K}$. Finally we show that our coercive field for $H_{\perp}$ peaks at $T_f \approx 12$\,K in Fig. \ref{MicroHall}(f) whereas Fig. 5(c) shows a monotonic increase of $H_c$ for $H_{\parallel}$. 12 K is the same temperature for which a freezing out of magnetic fluctuations is deduced from $\mu$SR. 

Micro-Hall measurements have previously been used to track Barkhausen jumps in micron-sized magnetic particles due to pinning/depinning of magnetic domain walls. \cite{Das10} There are no indications for Barkhausen jumps, which would be characteristic of conventional ferromagnetic domains, in the magnetization for Lu$_2$MnCoO$_6$.

\section{Discussion}

\begin{figure*}
\centering
\includegraphics[width=12 cm]{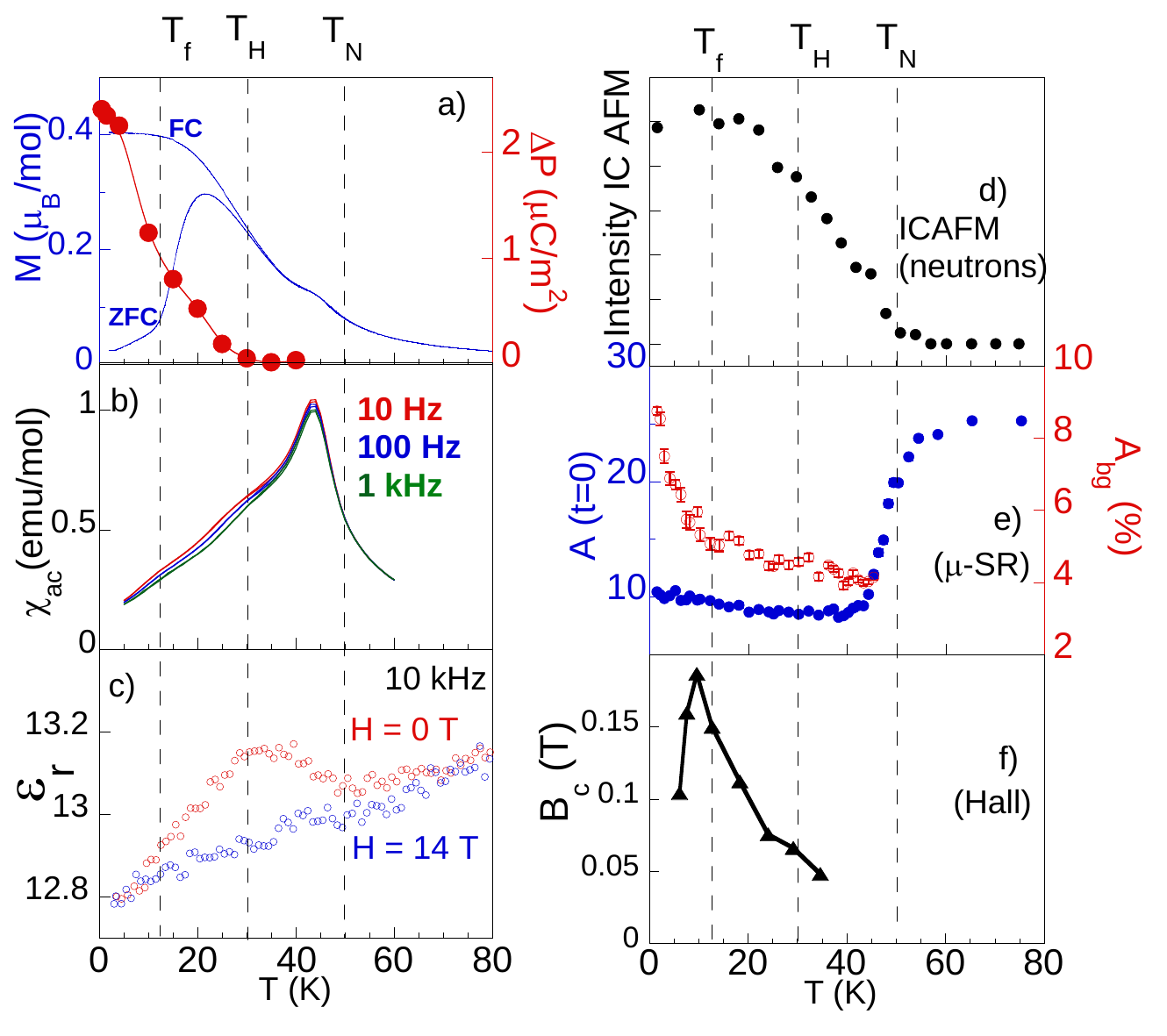}
\caption{$T$-dependence of different measured properties of Lu$_2$MnCoO$_6$, showing three temperature scales: $T_N$, $T_H$, and $T_f$. a) shows $M(T)$ after zero-field cooling (ZFC) and field-cooling (FC) in $\mu_0 H = 0.1$ T, as well as $\Delta P =  P(\mu_0H = 0~T) - P(\mu_0H = 15~T)$ \cite{YanezVilar11} b) ac magnetic susceptibility at 10 Hz, 100 Hz and 1 kHz, c) dielectric constant $\epsilon_r$ measured for $H || E$ with a static $\mu_0 H$ of 0 and 14 T and a small oscillating $H$ at 10 kHz, d) $T$-dependence of the ICAFM peak at 70.70$^{\circ}$ in the neutron scattering for $\mu_0 H = 0$, e) $T$-dependence of  $A(t=0)$ and $A_{bg}$  from $\mu$SR measurements, as described in the text, and f) $T$-dependence of the coercive magnetic field $B_c$ for $H_{\perp}$ from micro-Hall measurements. }
\label{TempSweep}
\end{figure*}

In Fig.~\ref{TempSweep} we summarize the temperature dependence of the different physical properties measured in this and previous work.  Fig.~\ref{TempSweep}a, b, and c show magnetization ($M$), electric polarization change ($P$), ac magnetic susceptibility ($\chi_{\textrm{ac}}$), and dielectric constant ($\epsilon_{\textrm{r}}$) data previously reported in Ya\~{n}ez-Vilar {\it et al} for comparison. \cite{YanezVilar11} 
In Fig.~\ref{TempSweep}(d)-(f) we show data from this work: the intensity of the ICAFM neutron diffraction peak from Fig.~\ref{neutron_Tdep}, and parameters extracted from the $\mu$SR and micro-Hall data as a function of $T$ on the same temperature scale. Three temperatures are indicated: the ordering temperature now identified as $T_N = 50 \pm 3$ K, the onset of magnetic and electric hysteresis at $T_H = 30$ K as well as a peak in the dielectric constant, and spin freezing with features in the magnetization at $T_{f} =12$ K.  

The neutron diffraction data in Fig.~\ref{TempSweep}(d) reveals that magnetic ordering onsets below $T_N = 50 \pm 3$ K.  The time scale for neutrons to interact with an ordered spin system is on the order of picoseconds, or THz. On the other hand, $\mu$SR spectra, although they show sharp features a few degrees below $T_N$, do not see any oscillations due to a combination of static disorder and dynamic disorder on MHz time scales. Thus the magnetic order below $T_N$ is static on THz time scales but strongly fluctuating on MHz time scales.
We define $T_N$ from neutron diffraction, but note that $T_N$ is a few degrees lower in the $\mu_{SR}$ and magnetization data, likely due to the slow dynamics in the system.

Moving on to the hysteresis temperature, $T_H \sim 30$ K, we see that at this temperature the zero-field cooled $M(T)$ curve separates from the field-cooled one, and a frequency dependence can be resolved in the ac magnetic susceptibility. A peak occurs in the 1 kHz dielectric constant. Hysteresis in $M(H)$ curves can be identified below this temperature in previously-measured magnetization data (Vibrating Sample Magnetometry and SQUID magnetometry), \cite{YanezVilar11, Lee14} and from micro-Hall magnetization measurements of the sintered monocrystallites in this work. 

An important point is that at $T_H$ we see no resolvable features in the neutron diffraction data nor in the $\mu$SR spectra. 

Neither the peaks nor the intensities of the neutron diffraction peaks show a feature at $T_H$. We conclude that at $T_H$ there is an onset of magnetic hysteresis and in dynamics on kHz time scales, but no change in the magnetic ordering on MHz or THz time scales.  As to the question of where electric polarization onsets, we note that $\Delta P$ in Fig.~\ref{TempSweep}(a) is a measurement not of the total electric polarization, but rather of the remanent electric polarization after the electric field is removed. Thus the onset of $\Delta P$ below $T_H$ indicates the onset of electric hysteresis, and occurs at the same temperature where the magnetization becomes hysteretic. On the other hand, an upturn can be resolved in the dielectric constant at the magnetic ordering temperature $T_N$ in Fig.~\ref{TempSweep}(c)  suggesting that ferroelectric order is induced at the magnetic ordering temperature $T_N$.

Finally below $T_f = 12$ K  the $\mu$SR shows the beginning of a freezing process on MHz scales of the fluctuating spins. At this temperature we observe a kink in the zero-field-cooled $M(T)$ curve and a peak in the coercive magnetic field $H_c$ from micro-Hall data. 


The strong fluctuations persisting to low temperatures and the 
$\uparrow\uparrow\downarrow\downarrow$ ordering of anisotropic moments along $c$-axis chains are features of ANNNI models. \cite{Bak82} The canonical ANNNI model assumes ferromagnetic nearest-neighbor exchange $J_0$ along all directions, with antiferromagnetic next-nearest-neighbor exchange $J_1$ along one direction.
For certain values of $J_0/J_1$, the $\uparrow \uparrow \downarrow \downarrow$ ground state is predicted, while $\uparrow \uparrow \downarrow$ are expected in other regions of phase space. For finite temperatures, long-wavelength modulations along the $J_1$ axis are predicted, with temperature-dependent sliding of the modulation length. 

For an ANNNI model on a discrete lattice, excitations in the form of domain wall solitons and antisolitons are predicted to occur, e.g. spin flip defects in the $\uparrow \uparrow \downarrow \downarrow$ ground state leading to $\uparrow \uparrow \uparrow \downarrow \downarrow$, etc. \cite{Bak82} These excited states are metastable, separated from the ground state by relatively high energy barriers such that relaxation into the ground state can occur on longer time scales than experiments. Such domain wall solitons were recently observed in Ca$_3$Co$_2$O$_6$ \cite{Agrestini08a,Agrestini08b,Agrestini11,Moyoshi11} with relaxation times on the order of days. An important feature of these domain wall solitons is that they can have diffuse modes, e.g. they can freely move up and down the chains, while preserving the underlying magnetic ordering. Thus, they provide a natural explanation for strong magnetic fluctuations observed in $\mu$SR.

Thus one explanation for the dynamics of Lu$_2$MnCoO$_6$ is in terms of excitations of an ANNNI model. Lu$_2$MnCoO$_6$ shows a departure from the standard ANNNI model in that two different magnetic species with different degrees of Ising anisotropy are present. Based on their orbital occupations, Co$^{2+}$ is Ising-like while Mn$^{4+}$ spins are more isotropic. \cite{Kim14}  Nevertheless, the principles of dynamic domain wall solitons forming along the $c$-axis should apply to a range of ANNNI-related systems. How the ANNNI ground state evolves in magnetic field to produce strong magnetic hysteresis, arising out of a net $M = 0$ ground state is the next question, and the subject of an ongoing study.

\section{Conclusion}

In conclusion, Lu$_2$MnCoO$_6$ is a multiferroic with the unusual and potentially useful property that an overall hysteretic net magnetization  (e.g. not an antiferromagnetic order parameter with $M = 0$ as occurs in many other multiferroics) couples to an equally hysteretic electric polarization, and this occurs in the bulk magnetic ordering. We identify three temperature scales in Lu$_2$MnCoO$_6$ corresponding to the onset of strongly fluctuating long-range order below $T_N = 50 \pm 3$ K, the onset of an overall net hysteretic magnetization and electric polarization below $T_H \sim 30$ K with no change in the ordering wave vectors, and finally the beginning of a spin freezing process of MHz-scale fluctuations, that also corresponds to features in the magnetization near 12 K. As a consequence, the magnetic and electric hysteresis onset at a lower temperature than the ordering temperature. Strong spin fluctuations are observed in $\mu$SR at all temperatures, consistent with the magnetic frustrated ANNNI scenario with dynamic domain wall solitons forming along the spin chains. Between the N\'{e}el temperature at 50 and 12 K, the material shows static magnetic order only on the time scale of neutrons (THz) and no static magnetic order on the time scale of muons (MHz). Freezing of the spins on muon time scales begins only below 12 K. On the other hand, while individual spins are experiencing strong fluctuations, an overall hysteretic net magnetization can be observed below 30 K, due to a conservation of net aligned spins in the presence of dynamic domain wall solitons. This material is a thus a candidate for multiferroic behavior resulting from the ANNNI model, where unusually strong coupled magnetic and electric hysteresis result from field-induced sliding domain wall solitons. 

\begin{acknowledgments}
Work at LANL was supported by the Laboratory-Directed Research and Development program (LDRD). The NHMFL Pulsed-Field Facility is funded by the U.S. National Science Foundation through Cooperative Grant No. DMR-1157490, the State of Florida, and the U.S. Department of Energy.  $\mu$SR measurements were carried out at  S$\mu$S, Paul Scherrer Institut, Switzerland and we are grateful to Alex Amato for technical assistance. That work is supported by EPSRC (UK). The Spanish authors are grateful for financial support from Ministerio de Economía y Competitividad (MINECO) (Spain) and EU under projects FEDER MAT2010-21342-C02.  Work at the Ames Laboratory was supported by the Department of Energy, Basic Energy Sciences, Division of Materials Sciences \& Engineering, under Contract No. DE-AC02-07CH11358.  Muon data presented in this paper will be made available via http://dx.doi.org/10.15128/qj72p712h.

\end{acknowledgments}

\bibliography{Lu2MnCoO6}


\clearpage

\end{document}


\title{Supplemental Information to: Coupled magnetic and electric hysteresis in the multiferroic double perovskite Lu$_2$MnCoO$_6$}

\author{Vivien S. Zapf,$^1$ B. G. Ueland,$^2$ Mark Laver,$^3$ Martin Lonsky,$^4$ Merlin Pohlit,$^4$ Jens M\"{u}ller,$^4$ Tom Lancaster,$^5$ Johannes S. M\"oller,$^{6,*}$ Stephen J. Blundell,$^6$ John Singleton,$^1$  Jorge Mira,$^7$ Susana Ya\~{n}ez-Vilar,$^8$ and Maria Antonia Se\~{n}ar\'{i}s-Rodr\'{i}guez$^8$}

\affiliation{$^1$National High Magnetic Field Laboratory, Los Alamos National Laboratory, Los Alamos, NM 87545, USA}
\affiliation{$^2$Ames Laboratory, U.S. DOE and Dept. of Physics, Iowa State University, Ames, Iowa 5011, USA}
\affiliation{$^3$School of Metallurgy and Materials, University of Birmingham, Edgbaston B15 2TT, UK} 
\affiliation{$^4$Institute of Physics, Goethe University, 60438 Frankfurt (M), Germany}
\affiliation{$^5$Durham University, Department of Physics South Road, Durham DH1 3LE, UK}
\affiliation{$^6$Oxford University, Department of Physics, Clarendon Laboratory, Oxford OX1 3PU, UK}
\affiliation{$^7$Departamento de F\'{i}sica Aplicada, Universidad de Santiago de Compostela, E-15782 Santiago de Compostela, Spain}
\affiliation{$^8$Dpto. Qu\'{i}mica Fundamental U. Coru\~{n}a, 15071 A Coru\~{n}a, Spain}
\affiliation{$^*$Now at the Laboratory for Solid State Physics, ETH Z\"urich, Z\"urich, Switzerland}

\maketitle

\section{Elastic neutron diffraction}

The integrated intensities of three peaks in the neutron diffraction are shown as a function of temperature in Fig. \ref{FigNeutSI}.  This data supports the notion that the magnetic order on quasi-static neutron timescales onsets below $T_N$ and does not show additional phase transitions at lower temperatures.

\begin{figure*}
\centering
\includegraphics[width=1.0\linewidth]{Lu2MnCoO6_SI_neutron.pdf}
\caption{The integrated intensities vs temperature for (a) the (110) Bragg peak, (b) the incommensurate antiferromagnetic Bragg peak (IC AFM), as defined in the body of the manuscript, and (c) the $(020)-\bm\tau$ and $(021)-\bm\tau$ Bragg peaks.  The $(020)-\bm\tau$ and $(021)-\bm\tau$ Bragg peaks overlap for the experimental resolution.  The dashed lines are guides to the eye.  $\bm\tau = (0.0223(8), 0.0098(7), 0.5)$.  The integrated intensities were determined from fits to Gaussian lineshapes.  Note that for (110), the measured data did not completely span the Bragg peak for each temperature.  Hence, to determine the integrated intensities shown in (a), we fixed the values for the center and the full width at half maximum in the fits to the values determined at $T=1.6$ K. }

\label{FigNeutSI}
\end{figure*}

\section{Muon spin relaxation}

Example  $\mu$SR spectra measured at three temperatures are shown in
Fig.~\ref{muSR_SI}. The spectra at all temperatures show monotonic
relaxation described by one or more exponential functions. This is
typical of relaxation caused by dynamic fluctuations of the magnetic
field distribution experienced by the muon ensemble. 
The behavior may be separated into three regimes. Above $T_{\mathrm{N}}=50$~K we
observe relaxation with the full relaxing fraction of
asymmetry. For $12 \leq T \leq 50$~K the initial asymmetry is reduced,
but the spectra continue to relax to the same baseline. For $T<12$~K
the apparent baseline increases and the relaxation rate is reduced. 

\begin{figure}
\begin{center}
\includegraphics[width=8 cm]{muSR_SI.pdf}
\caption{Zero field $\mu^{+}$SR spectra as a function of time at $T =$ 1.5, 30, and 50 K}. \label{muSR_SI}
\end{center}
\end{figure}

In order to parametrize the behavior of the system across the
temperature regime, the spectra were fitted to two relaxation
functions. For $T>T_{\mathrm{N}}$ the spectra were found to be well
described by two exponential functions:
\begin{equation}
A(t) = A_{\lambda}e^{-\lambda t}+A_{\Lambda}e^{-\Lambda t}+A_{\mathrm{bg}},
\label{tgttn}
\end{equation}
where $\lambda$ and $\Lambda$ are relaxation rates. The two
exponential functions imply the existence of two magnetically distinct
classes of muon site. 
We find that at temperatures well above 50~K, the amplitudes take constant values 
$A_{\lambda}=13.5$~\%, $A_{\Lambda}=7.7$\%
and the baseline is $A_{\mathrm{bg}}=3.9$\%. 
We also find that the
relaxation rate $\Lambda$ takes values far larger than $\lambda$, 
although the ratio of the two remains in a roughly fixed
proportion, implying that they are tracking the same physics and only
differ due to the position of the respective muon site in the unit
cell. 
For $T<T_{\mathrm{N}}$ we are only able to resolve a single
exponential relaxation rate and data are fitted to
\begin{equation}
A(t) = A_{3}e^{-\lambda t}+A_{\mathrm{bg}}. 
\label{tlttn}
\end{equation}

The baseline amplitude $A_{\mathrm{bg}}$ is not constant below
$T_{\mathrm{N}}$, but is seen to increase significantly below
$T_{\mathrm{f}}=12$~K (main text).

The evolution of the initial asymmetry $A_\lambda(t=0)$ with temperature is shown
in the main text. The sharp drop that is apparent around
$T_{\mathrm{N}}=50$~K is typical of a system undergoing a transition
to a regime showing some form of magnetic order. The large, slowly
fluctuating electronic moments that develop below $T_{\mathrm{N}}$ cause muon spins to evolve
very rapidly upon implantation. The average polarization of these muons is dephased
within the first time bin of the measurement and only a residual
relaxation is observed. However, the lack of oscillations in the
asymmetry suggests that the transition is not one to a regime of
quasistatic long range magnetic order (LRO) on the muon (microsecond)
timescale. In this regime we expect the relaxation to
vary as $\lambda \propto \gamma_{\mu}^{2}\langle (B-\langle
B\rangle)^{2}\rangle \tau$, where $\tau$ is the relaxation time, $B$ 
is the local magnetic field at the muon site(s)
and $\gamma_{\mu}$ is the muon gyromagnetic ratio. 
The large magnitude of  the relaxation we observe points to the
possible combination of a broad distribution of local magnetic fields
and the presence of dynamic fluctuations. 
The presence of dynamics is confirmed by the fact that the spectra in
the region $12 \leq T \leq 50$~K relax to the
same
baseline value as for $T>T_{\mathrm{N}}$. For a powder sample we would
expect that 1/3 of muons should have spins
initially polarized along the direction of the  magnetic
field. In a {\it static} magnetic state these would not be depolarized and lead to an increase in the
apparent baseline of the relaxation compared to that found for
$T>T_{\mathrm{N}}$.  
The state just below $T\approx 50$~K
therefore appears to be one showing large moments with some degree of disorder, which are
dynamically fluctuating on the muon time-scale. 

The increase in the baseline amplitude $A_{\mathrm{bg}}$ below 12~K is
strongly suggestive of a freezing out of these dynamics in the local
magnetic field distribution experienced by the muons. In this case the
2/3 of the muon spins initially oriented perpendicular to the local
magnetic field direction will still be dephased by the variation in
the static fields across the muon ensemble, but those 1/3 of muon
with spins directed parallel to the local magnetic fields will not be
dephased in the absence of dynamics.
It is worth noting that this behavior was also observed
in the low temperature behaviour of Ca$_{3}$Co$_{2-x}$Mn$_{x}$O$_{6}$
for $x=0.95$ \cite{Lancaster09} and for other values of $x$. \cite{tom2}

\section{Micro-Hall data}

\begin{figure}
\begin{center}
\includegraphics[width=8 cm]{MicroHall_SI1.png}
\caption{
(a) A sample with irregular shape has been positioned
on a 10 x 10 $\mu$m$^2$ Hall cross. The blue square corresponds
to the active area of the cross. Another sample
consisting of only $\sim$4 particles was positioned on the next-nearest
cross, but due to the low filling factor the signal-to-noise
ratio is too low to allow for sensitive measurements.
Inset: Lu$_2$MnCoO$_6$ powder sample consisting of sintered particles
with an average diameter 1-2 $\mu$m. (b) Enlargement
showing that the sample consists of $\sim$40 particles. Directions
of the externally applied field with respect to the sensor are
indicated.
} \label{MicroHall_SI1}
\end{center}
\end{figure}

Micro-Hall measurements were performed by placing a small number of sintered samples onto a Hall cross. The stray field emanating from the sample, perpendicular to the Hall plane is recorded as $\langle B_z \rangle$.
When measuring a very small signal on top of a large background as it is the case for the perpendicular field geometry,
so-called gradiometry measurements are performed, where the ordinary linear background is cancelled directly
during the measurement by applying an amplitude and phase-adjusted current in opposite direction to an
empty reference cross. However, a small nonlinear background, which is related to the ballistic nature of electron
transport in the sensor and subtle differences between different Hall crosses, remains. This strongly temperature-dependent
nonlinear background can be subtracted subsequently from the data in order to obtain the net hysteresis
loop of the magnetic sample. Measurements have been performed in a temperature range from 50K down
to 0.3 K. In addition, for the perpendicular orientation of the external field with respect to the sensor, quantum oscillations
and the quantum Hall effect in a  2-dimension electron gas may complicate the background subtraction at low temperatures.
From a Lu$_2$MnCoO$_6$ powder sample shown in the inset of Fig.~\ref{MicroHall_SI1}(a) an ensemble of microparticles was extracted
and positioned on a Hall sensor with a micromanipulator. This sample shows an irregular shape and consists of
roughly 40 sintered monocrystalline grains, nicely showing the faceted morphology that typically is displayed by
single crystals of perovskites. Fig.~\ref{MicroHall_SI1}(a) also shows another sample of smaller size placed on the next-nearest
cross. Whereas the 'filling factor' of the smaller sample was too low to allow for sensitive-enough measurements,
the enlarged view of the larger sample in Fig.~\ref{MicroHall_SI1}(b) clarifies an almost optimal ratio of sample and cross size,
which results in a large signal-to-noise ratio. Measurements were taken for $H$ parallel  ($H_{\parallel}$) and perpendicular ($H_{\perp}$) to the sensor plane. Comparing our data to single crystals, we conclude that the c-axis of the crystallites is predominantly perpendicular to the sensor plane. \cite{Lee14}

The magnetic hysteresis loops are shown in the main text. Exchange bias is observed for $H_{\parallel}$ but not $H_{\perp}$. This could be due to the ferrimagnetic nature of the magnetic order containing both Co$^{2+}$ and Mn$^{4+}$ spins.  For $H_{\perp}$ in Fig.~\ref{MicroHall_SI2} we find that the remanent magnetization depends on the applied saturation field when $T < T_f$. This does not occur for $H_{\parallel}$. Example data is shown at 10 K. In particular, a distinct dependence occurs for low saturation fields, whereas it becomes less pronounced for higher fields.   

Finally, we show a magnetic hysteresis loop at 0.3 K for $H_{\parallel}$, that shows a sharp jump in the magnetization at a 2 T coercive magnetic field. This is consistent with the similar sharp jump observed below 2 K in polycrystalline data. \cite{YanezVilar11}

\begin{figure}
\begin{center}
\includegraphics[width=8 cm]{MicroHall_SI2.pdf}
\caption{
Magnetic hysteresis loops (raw data) for $H_{\perp}$ at $T = 10$ K showing a strong dependence
of magnetic remanence on the maximum applied magnetic field of the hysteresis loop.
} \label{MicroHall_SI2}
\end{center}
\end{figure}

\begin{figure}
\begin{center}
\includegraphics[width=8 cm]{MicroHall_SI3.png}

\caption{Magnetic hysteresis loop for $H_{\parallel}$ at 0.3 K.}
 \label{MicroHall_SI3}
\end{center}
\end{figure}

\bibliography{Lu2MnCoO6}